# Softwaretechnische Absicherung intelligenter Systeme im Fahrzeug

# Software Engineering Methods for Quality Management of Intelligent Automotive Systems

Prof. Dr. **B. Rumpe**, **C. Berger**, **H. Krahn**,
Software Systems Engineering, TU Braunschweig

**Kurzfassung**

Dieser Artikel skizziert, welche Software-Entwurfstechniken heute zum Einsatz kommen können, um intelligente, Software-lastige Systeme im Fahrzeug abzusichern. Dabei spielt zunächst das Qualitätsmanagement durch Software-technische Maßnahmen eine zentrale Rolle. Architektur- und Entwurfmuster für die Software-technische Absicherung von Komponenten werden ergänzt um Test-Konzepte zur Validierung von Spezifikationen und der Robustheit der Implementierung. Architekturen und Entwurfs-Muster beschreiben erprobte und damit konsolidierte Lösungen für bestimmte Problemklassen wie etwa Zuverlässigkeit oder effiziente Ausführung.

**Abstract**

This article describes software engineering techniques to be used in order to ensure the necessary quality of intelligent and therefore massive software-based systems in vehicles. Quality assurance for intelligent software is achieved through a bundle of modern software engineering methods. Architecture and design patterns for securing the software components are supplemented by test concepts and frameworks for validation and checks of robustness of the implementation. These patterns describe established and therefore consolidated solutions for certain problems as for instance reliability or efficient execution.

## 1. Einleitung

Ein deutlicher Teil der Innovationen im Automobilbereich findet bei den Fahrerassistenzsystemen statt. Fahrerassistenz erfordert ein konsequentes und integriertes Management der Entwicklung von mechanischen, elektronischen und rasant komplexer werdenden erkennenden, vorausschauenden und steuernden Software-Funktionen. Aktuelle Trends zeigen, dass intelligente Systeme neben einer Unterstützung im Komfortbereich vor allem im Bereich der Fahrerassistenz und der Sicherheit aller Verkehrsteilnehmer Einsatz finden. Der „intelligente Beifahrer" warnt bei unangemessenem Fahrverhalten, Müdigkeit und greift ggf. durch Bremsen oder Lenken aktiv in die Fahrzeugsteuerung ein.

Der konsequente Aufbau und die Weiterentwicklung heutiger Fahrerassistenzsysteme erfordert die Integration komplexer, oft stark voneinander abhängiger Software-Komponenten. Zusammen bilden sie den „intelligenten Beifahrer", beinhalten teilweise echte Intelligenz und sind nach gängigen Hardware-Architekturen immer noch auf unterschiedlichen Steuergeräten im Fahrzeug realisiert. Der Wettbewerbsvorteil durch die aus der Software-Branche bekannten kurzen Entwicklungszeiten von der Idee zur Umsetzung im Fahrzeug einerseits und die teilweise komplexe, erlernte und damit schwer beherrschbare Intelligenz andererseits verhindern den Einsatz fundierter Methoden zur Verifikation der implementierten Funktionalität. Insbesondere die Verwendung von probabilistischen oder lernenden Algorithmen, die Prinzipien der künstlichen Intelligenz in ein Fahrzeug integrieren, sind aufgrund ihrer Funktionsweise nur aufwändig mit traditionellen Maßnahmen zur Qualitätssicherung zu behandeln, da sie unstetiges Verhalten und Modi-Wechsel beinhalten, die nicht notwendigerweise vorhersagbar sind. Außerdem nehmen traditionelle Maßnahmen relativ lange Zeit in Anspruch, denn Software und damit Software-Entwicklung ist auch aufgrund fehlender bzw. für Autos zu teurer Abstraktionsmechanismen üblicherweise von der zu steuernden Hardware (Sensorik, etc.) abhängig und wird daher erst spät im Gesamtentwicklungsprozess integriert.

Die rasant angestiegene Komplexität sowie die zu erwartende intensive Vernetzung der Software-Funktionen, die von verschiedenen Herstellern/Zulieferern entwickelt werden und verteilt über Steuergeräte und oft sogar Bussysteme ihre Funktionalität erbringen, erfordern neue Konzepte und Methoden zur Entwicklung und Qualitätssicherung. Die Absicherung von Funktionen beginnt im Softwareentwicklungsprozess durch Anwendung qualitätssichernder Maßnahmen und nutzt gleichzeitig Architektur-Prinzipien zur Laufzeit-Absicherung. Neben physisch sichtbaren Architekturen (Auto-Aufbau und Busstruktur) hat sich bei der Softwareentwicklung eine weitere Architekturform gebildet: Die Softwarearchitektur. Das Konzept der Softwarearchitektur steht (relativ) orthogonal zur Bus- und Steuergerätestruktur, bietet aber ähnliche Charakteristika. So erlaubt es den Softwareentwicklungsprozess zu strukturieren und bietet Möglichkeiten zur Weiterentwicklung und Wiederverwendung einzelner Software-Komponenten und –Funktionen. Erprobte Musterarchitekturen können zur Lösung immer wiederkehrender Probleme eingesetzt werden. Solche Muster eignen sich, um Funktionen geeignet auf Steuergeräte zu partitionieren oder eine softwaretechnische Absicherung sicherheitskritischer Funktionen zu realisieren. Dadurch können Entwicklungszeiten verkürzt, die Qualitätssicherung verbessert, die Wiederverwendung innerhalb einer Produktlinie erhöht und durch den Einsatz aktueller

Stand-der-Technik-Methoden des Software Engineerings auch den unausweichlichen Haftungsfragen im Bereich Fahrerassistenzsysteme aktiv entgegnet werden.

Fahrerassistenzsysteme werden im Allgemeinen als „intelligent" oder zumindest „semi-intelligent" betrachtet. Die Informatik hat mittlerweile eine Reihe von Arten der Intelligenz exploriert [7]. Eine einfache Klassifikation von Intelligenz unterscheidet zwischen lernenden Algorithmen (und damit „echter") Intelligenz und statischen, unveränderbaren Algorithmen. Eine weitere Unterscheidung kann getroffen werden zwischen Lernen in einer Entwicklungsphase und einem adaptivem Lernen während der Benutzung. Die Absicherung adaptiver Systeme stellt eine besondere, nachfolgend diskutierte Herausforderung dar.

## 2. Methoden der Software-Architektur

Der Begriff der Software-Architektur hat sich in den letzten Jahren als zentrales Strukturierungsmittel für die Softwareentwicklung etabliert [7]. Ähnlich der Architektur von Gebäuden, der Komponenten in komplexen Maschinen oder elektronischen Schaltungen dient sie zur Zerlegung eines komplexen Problems in einzelne Komponenten, die parallel entwickelt oder eingekauft und ggf. unabhängig voneinander eingesetzt werden können. Mittlerweile bildet sich sogar das Berufsbild des „Software-Architekten" als zentrale technische Schaltstelle für Projektentwicklung und –management heraus. Mit AUTOSAR [15] wird auch im Bereich des Automobilbaus langsam Architektur als Planungsgröße, die nicht nur durch Hardwarestruktur getrieben ist, integriert. Weil jedoch Software immateriell ist, ist notwendigerweise auch Software-Architektur immateriell und damit für Nicht-Software-Entwickler schwer erfahrbar. Architekturen werden gerne in Form von Kästchen und dazwischen liegenden Pfeilen skizziert, die aber je nach Anschauungskontext sehr verschiedene Bedeutung haben können, denn es gibt auf die Architektur eines Software-Systems mehrere nur partiell voneinander abhängige Sichten, die der Softwarearchitekt beherrschen muss. Dazu gehören:

- Die Aufgabensicht: Welche Aufgaben sind zu lösen? Welche funktionalen Anforderungen zu erfüllen?
- Die funktionale Sicht: Welche Software-Funktionen existieren und wie sind ihre Abhängigkeiten?
- Die Software-Komponentensicht: Welche Komponenten (Module) realisieren welche Funktionen und welche Schnittstellen bestehen zwischen ihnen?
- Die Schichtenbildung: Von der obersten Applikationsebene bis zur untersten Schicht des Betriebssystems mit ihren Kommunikations-, und Speicherprimitiva.

- Die Prozesssicht: Welche parallelen Prozesse existieren? Sind diese dynamisch startbar?
- Die Prozessorsicht (Hardware): Welche Prozessoren sind im System verbaut? Welche Prozesse laufen auf welchen Prozessoren und sind diese verlagerbar?
- Die Bussicht: Welche Busse existieren und sind wie mit Prozessoren und Gateways verbunden?
- Die Software-Paketsicht: Welche Software-Module sind in welche Pakete eingebunden? Welche statischen Abhängigkeiten (Imports, Includes) existieren?
- Die Entwickler-Sicht: Wer übernimmt Verantwortung für welche Module, Prozesse bzw. andere Einheiten der modellierten Softwarearchitektur?

Zwischen diesen Sichten sind Abbildungen, oft auch als „Partitionierung" bezeichnet, zu definieren, welche die Konsistenz der einzelnen Sichten widerspiegeln. Beispielsweise erfüllen meist mehrere Funktionen eine Aufgabe. Ein Prozess kann mehrere Funktionen realisieren, aber eine Funktion kann redundant auch von mehreren Prozessen erfüllbar sein, um etwa Ausfälle abzusichern. Weitere Architektursichten sind unter Umständen sinnvoll, wenn etwa hohe dynamische Anteile vorhanden sind oder Diagnose als integraler Bestandteil des Systems relevant sind.

Die Anforderungen an ein Software-System werden oftmals in funktionale und nicht-funktionale Anforderungen unterteilt. Für Software-Systeme relevante nicht-funktionale Anforderungen werden dort als Qualitätsattribute bezeichnet. Darunter fallen im Allgemeinen Bedingungen über Verfügbarkeit, Modifizierbarkeit, Performanz, Sicherheit im Sinne von Safety (sowie auch der im Automobilkontext derzeit weniger relevanten Security), Testbarkeit und Verwendbarkeit ([2], [9]). Viele dieser Qualitätsattribute sind im Wesentlichen abhängig von der Wahl der richtigen Software-Architektur. Um erprobte Architekturen wieder verwenden zu können wurde in [2] eine Klassifikation von Architekturtaktiken entworfen. Unter einer Taktik werden einzelne Entwurfsentscheidungen verstanden, um bestimmte Qualitätsattribute in Systemen zu realisieren. Eine Taktik ist dabei eine grundlegende Entscheidung, ein geeignetes Muster einzusetzen um so seine Vorteile und Fähigkeiten zu nutzen.

Muster spielen in der Entwicklung von Software-Architekturen zunehmend eine entscheidende Rolle, da sie von konkreten Systemen abstrahierte Lösungen für wiederkehrende Herausforderungen darstellen. Mustersammlungen, wie zum Beispiel in [1], [5], [6], [7] und [8] enthalten, stellen einen zentralen Erfahrungsschatz des Software Engineering dar und erlauben die Wiederverwendung von erprobten Lösungen durch Übernahme von Konzeptionen zwischen verschiedenen Projekten und Domänen. Sie bilden daher ein zentrales Element im Portfolio des Software Engineering, das auch Software-

Bibliotheken, Analyse- und Generierungswerkzeuge sowie Verwaltungs- und Managementwerkzeuge kennt und alles in geeigneten Entwicklungsprozesse integriert. Ein Muster löst typischerweise eine Klasse von verwandten Problemen und kann daher in mehreren Taktiken zum Einsatz kommen. Relevant dabei ist, dass mit dem Muster sowohl die lösbare Problemklasse beschrieben wird, als auch Vorteile und potentielle negative Auswirkungen auf verschiedene Qualitätsattribute explizit diskutiert werden. So lässt sich zum Beispiel mit zusätzlichen softwaretechnischen Sicherungsmechanismen eine höhere Systemverfügbarkeit erreichen, die aber mit einer etwas schlechteren Antwortzeit des Systems erkauft sein kann. Aufgabe des Software-Architekturentwurfs ist damit auch die Beurteilung der Auswirkungen verschiedener Entwurfsentscheidungen auf das zu entwickelnde System.

## 2.1 Fehlererkennung und -vermeidung

Während eines Systembetriebs können auf verschiedenen Stufen der Verarbeitung von Informationen Fehler auftreten. Die Ursachen für solche Fehler sind vielfältig und können fehlerhafte Sensoren, gestörte Übertragungen, Hardware-Defekte in Steuergeräten, Fehler anderer Systeme, aber auch fehlerhafte Software sein. Unabhängig von der Ursache kann dabei zwischen einem Defekt und einem Ausfall unterschieden werden. Ein Defekt bezeichnet ein unerwünschtes Verhalten eines Subsystems, wohingegen ein Ausfall den Fehler eines Systems bezeichnet, der nach außen beobachtbar ist. Somit kann ein nicht erkannter Defekt zu einem Ausfall des Gesamtsystems werden. Diese Unterscheidung zwischen Fehlerursache, Fehlerwirkung und tatsächlichem Versagen des Systems ist bei komplex vernetzten Systemen besonders relevant, da vielfältige Wirkketten möglich und damit Absicherungsmaßnahmen nötig sind.

Im Zusammenhang mit Fehlern eines Systems sind die Qualitätsattribute Verfügbarkeit mit der Erkennung und Behebung von Ausfällen und Sicherheit mit den finanziellen und lebensbedrohenden Auswirkungen eines Ausfalls betroffen. Daher unterscheiden sich diese beiden Qualitätsattribute in ihren Messmethoden, während die grundlegenden Taktiken zur Vermeidung von Fehlern häufig auf beide anwendbar sind.

Dabei ist wichtig, dass die Qualitätsattribute auf verschiedenen Systemebenen, Architektursichten und Integrationsstufen betrachtet werden. Ist ein Teilsystem in ein Gesamtsystem integriert, kann ein Ausfall des Teilsystems immer noch durch eine Strategie des Gesamtsystems behoben werden, ohne dass dies zu einem Ausfall des Gesamtsystems führt. Dieses Prinzip ist aus der mechanischen und elektronischen Welt hinlänglich bekannt und kann auf Softwarefunktionen entsprechend übertragen werden. Dabei allerdings zu berücksichtigen, dass Software keinem Verschleiß unterliegt und damit die Ausfallrisiken im Prinzip von der Lebensdauer des Systems unabhängig sind. Aus zwei Gründen sind jedoch bei intelligenter Software Ausfallrisiken ebenfalls zeitabhängig. Zum einen stellen sich Nutzer auf Softwarefehler ein, indem sie bestimmte Handlungsketten zu vermeiden lernen und so Ausfallzeiten reduzieren. Zum anderen kann adaptive, also zur Laufzeit lernende Software ebenfalls ihr Verhalten ändern. Dies kann sich positiv auswirken, indem sie Fehler vermeiden lernt. Da Fehler in unserem Kontext Ausfälle essentieller Fahrfunktionen sind, sollten jedoch Fehler in diesem Kontext nicht auftreten, so dass ein Lernen aus Fehlern in diesem Kontext nicht wirklich auftreten kann und Adaption damit höchstens negative Auswirkungen auf die Zuverlässigkeit haben kann.

Die Ausfallsicherheit von Softwarefunktionen wird weiterhin stark durch die Integration vieler Funktionen auf einem Steuergerät und einem Prozessor beeinflusst. Ausführungszeiten und Speicherbedarf hängen stark voneinander ab, weil aus Kostengründen (noch) darauf verzichtet wird Abschottungsmechanismen einzubauen, die dies verhindern. Es ist ohne weiteres möglich, auf einem Prozessor gleichzeitig eine Funktion und deren Überwachung zu integrieren. Diese Technik erfordert möglicherweise einen stärkeren Prozessor und mehr Speicher, vermeidet aber zusätzliche überwachende Steuergeräte oder gar eine hardware-basierte Fail-Safe-Konstruktion.

## 2.2 Absicherung intelligenter Systeme

Einem intelligenten System wird üblicherweise eine hohe Komplexität zugesprochen. Diese Komplexität ist besonders schwer abzusichern, wenn es sich um das Zusammenwirken mehrerer verteilter Funktionen handelt, von denen einige eine gewisse Intelligenz beinhalten. Dennoch kann mit den üblichen Techniken des Software Engineering meist eine gute Qualitätssicherung erreicht werden. Besonders schwierig wird die Situation, wenn wie eingangs erwähnt die Intelligenz zum Teil darin besteht, dass die Software zur Laufzeit weiter lernt und damit ihr Verhalten modifiziert. Zur Absicherung solcher Software werden besondere Methoden benötigt, die teilweise heute noch in Entwicklung sind. Die teilweise spezifischen Herausforderungen sind hier:

- Eine Funktion wird nicht mehr korrekt erbracht, weil das System etwas Unvorhergesehenes oder Falsches gelernt hat. Dann kann das System falsche Schlüsse ziehen.
- Die Reaktionszeit wird zu lang und das System reagiert zwar korrekt aber zu spät.
- Die Softwarekomponente „stürzt ab".

Adaptive, lernende Software kann als parametrisierte Software verstanden werden, bei der sich die Parameter mit der Zeit modifizieren. Werden nur einfache Parameter (Zahlen) gelernt, so kann die Software mit relativ handelsüblichen Mitteln über die Bandbreite der Parameter validiert werden. Darüber hinaus ist es bei einfachen Parametern typischerweise möglich, Wertebereiche anzugeben, die auch bei adaptivem Verhalten nicht verlassen werden dürfen. Solche Wertebereichen können typischerweise geprüft und gegebenenfalls reinitialisiert werden. Dazu müssen die Wertebereichsgrenzen geklärt sein und eine Überwachungsfunktion zur Verfügung stehen. Die Adaption solcher einfachen Parameter beeinflusst meist nicht die Reaktionszeit oder Absturzwahrscheinlichkeit einer Funktion.

Wenn ein Parameter jedoch eine komplexe Datenstruktur erlaubt, dann können sich durch dynamische Änderungen eine Reihe von Problemen ergeben. Zum einen kann der Speicherbedarf und mit ihr die Reaktionszeit unangenehm wachsen. Dies kann dazu führen, dass das System nicht mehr rechtzeitig reagiert oder aufgrund nicht idealer Scheduling-Strategien andere Softwarefunktionen betroffen sind. Schlimmstenfalls reicht der Speicher nicht mehr aus und die Softwarekomponente „stürzt tatsächlich ab".

Zur Absicherung auf diese Weise parametrisierter Software ist es notwendig, eine Vielzahl lernbarer Situationen zu validieren. Es ist unumgänglich, Obergrenzen für lernbare Fakten zu entwickeln, gegebenenfalls eine Vergessensfunktion einzubauen oder die Lernfähigkeit zu beschränken. Fallbasiertes Lernen neigt sehr leicht dazu in diese Probleme zu laufen. Als sehr kritisch anzusehen ist, dass normalerweise weder ein Wertebereich noch eine Prüf-Funktionalität existiert, die die Parameter einer solchen Softwarekomponente auf ihre Korrektheit prüft und daher eine Reinitialisierung nicht automatisierbar ist. Entsprechend aufwändig ist es auch, das Verhalten einer solchen intelligenten Komponente zu prüfen um gegebenenfalls einzugreifen.

Die Sicherheit eines Systems wird typischerweise durch das Risiko in Form eines Produkts zwischen Ausfallwahrscheinlichkeit und Ausfallkosten charakterisiert. Daher behandelt Risikomanagement sowohl die Reduktion der Ausfallswahrscheinlichkeit als auch durch Milderung der Auswirkungen eines Ausfalls erreichen. In [2] und [9] werden Taktiken zur Erhöhung der Verfügbarkeit und Sicherheit in die folgenden grundlegenden Kategorien unterteilt:

- Erkennen des Fehlers durch geeignete Überwachung
- Vermeidung des Auftretens von Ausfällen durch Sicherungsmechanismen
- Abmilderung der Auswirkungen eines Ausfalls durch
    - Reparatur der beteiligten Software-Komponenten

o   Austausch der beteiligten Software-Komponenten durch Einführung neu initialisierte Komponenten
        o   Redundanz

Im Folgenden werden zwei aufeinander aufbauende Architekturmuster vorgestellt, welche die Verfügbarkeit eines Systems erhöhen, indem sie bestimmte Fehler erkennen und alternatives Verhalten anbieten.

**2.2 Liveliness-Prüfung durch das Ping-Verfahren**

Beim Ping-Verfahren [2] schickt eine zentrale Sicherungseinheit periodisch ein Signal an die von ihr zu überwachenden Komponenten. Diese beantworten die Anfrage innerhalb einer vorgegebenen Zeitschranke mit einer definierten Nachricht. Bleibt eine dieser Antworten aus, startet die Sicherungseinheit die Komponente neu bzw. reinitialisiert die Funktion und wechselt gegebenenfalls vergleichbar zu einem Watchdog während dieses Neustarts in einen Fehlerzustand, in dem eine alternative Komponente die Funktion übernimmt. Varianten dieses Verfahrens verwenden eine kaskadierte Kette von Komponenten, um den Kommunikationsaufwand des Verfahrens zu minimieren und die Austauschbarkeit von Komponenten zu erleichtern. Software- und Sicherungsfunktionen müssen nicht notwendigerweise auf verschiedenen Prozessoren lokalisiert sein, wenn die Abschottung der Prozesse innerhalb eines Prozessors ausreichend ist.

Das Ping-Verfahren birgt das Problem, dass sich nur Verzögerung und Ausfall einer Software-Komponente detektieren lassen, wohingegen ein funktionell fehlerhaftes Verhalten unerkannt bleibt. Um dies zu erkennen kann eine Watchdog-Funktion eingesetzt werden, die das Ausgabeverhalten auf Basis des Eingabeverhaltens prüft und fehlerhaftes Verhalten entdeckt. Wie bereits besprochen ist jedoch die Erkennbarkeit fehlerhaften Verhaltens bei intelligenten lernfähigen Systemen begrenzt. Es ist einerseits schwer vorherzusagen, wann ein System das Falsche lernt und andererseits die Bandbreite erlaubten Verhaltens gegebenenfalls deutlich höher als die individuell als angenehm empfundene Bandbreite des vom Nutzer erwünschten Verhaltens.

Das Ping-Architekturmuster lässt sich jedoch in einer Form erweitern, sodass von der Sicherungseinheit regelmäßig (oder in unkritischen Situationen) Testdaten an die Komponente gesendet werden. Die tatsächliche Reaktion des Systems wird mit einer fest in der Sicherungseinheit implementierten gewünschten Bandbreite an korrekten Antworten verglichen. Die Sicherungseinheit überprüft damit das korrekte Verhalten der Komponente und filtert die Testantworten entsprechend aus dem laufenden Betrieb aus. Dieses Verfahren eignet sich besonders gut für die Absicherung intelligenter Systeme, die ein eigenständiges reaktives Verhalten aufweisen, ist aber in laufenden Fahrbetrieb mit großer Vorsicht einzusetzen. Durch diese Sicherung kann quasi zur Laufzeit des Systems überprüft werden, ob die Reaktionen in fest definierten Notsituationen noch adäquat wären. Eine Übersicht über das Muster ergibt sich aus Abbildung 1. Zusätzlich ist zu sichern, dass die Testdaten keinen schädlichen Einfluss auf das Adaptionsverhalten des Systems nehmen, indem etwa die Lernparameter der intelligenten Softwarekomponente im Testfall nicht angepasst werden.

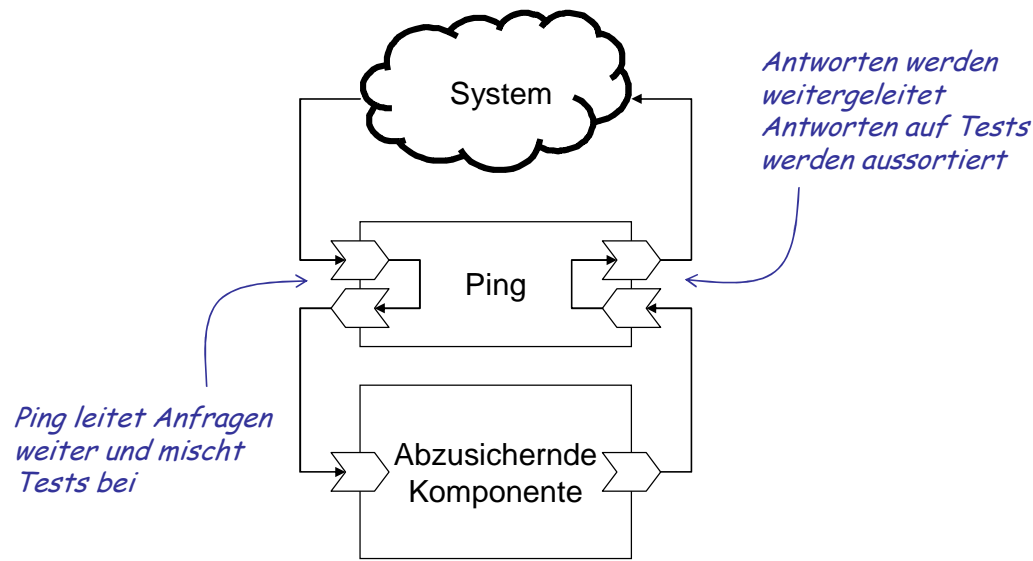

**Abbildung 1:** Darstellung der Einbindung des erweiterten Ping-Musters in ein System

## 2.3 Redundante Realisierungen

Bei aktiver Redundanz rechnen stets zwei Softwarekomponenten parallel. Bei Bedarf kann zeitnah auf die Ersatzkomponente umgeschaltet werden. Bei passiver Redundanz kann nach dem Ausfall bzw. nach dem Erkennen von fehlerhaftem Verhalten eine Ersatzkomponente bzw. ein neuer Prozess mit reinitialisierter intelligenter Softwarekomponente gestartet werden, die nach einer Synchronisationsphase die Funktionalität des Systems übernimmt. Um Ausfälle von Hardware-Komponenten zu bewältigen können die Softwarefunktionen auf verschiedene Prozessoren verteilt werden. Aber auch innerhalb eines Prozessors ist solche Redundanz sinnvoll, wenn zwei verschiedene Software-Implementierungen vorliegen und der Absturz einer Software-Komponente nicht den des gesamten Betriebssystems impliziert. Eine Standardtechnik hier besteht in der Einführung eines Redundanzsystems, das deutlich weniger komplex als das eigentliche Fahrerassistenzsystem ist und nicht die gleichen Anforderungen an den Komfort gewährleistet, aber die grundlegende Funktionalität sichert. Durch die geringen Anforderungen in Bezug auf Komfort lassen sich solche Softwarekomponenten meistens jedoch weniger aufwändig und kostengünstiger realisieren.

Damit lassen sich grundlegende Maßnahmen in Bezug auf Sicherheit und Qualität leichter sicherstellen als in Systemen, die adaptive Intelligenz beinhalten, die aufgrund der Art der Entscheidungsfindung schwerer zu testen sind. Das Redundanzsystem muss getrennt entworfen werden und sollte keine adaptiven Anteile beinhalten, damit durch den unterschiedlichen konzeptuellen Entwurfs davon ausgegangen werden kann, dass dieselben Konzeptionsfehler nicht in beiden Systemen auftreten und daher ein Zugewinn an Sicherheit und Verfügbarkeit zu erwarten ist. Eine vereinfachte und weniger sichere Form ein redundantes System einzuführen besteht darin, die lernende Software-Komponente ein zweites Mal, aber in permanent (oder häufig) initialisiertem Zustand laufen zu lassen. Natürlich kann durch die Einführung dreier redundanter Komponenten sogar eine Auswahl auf Basis eines Mehrheitsvotums vorgenommen werden.

### 2.4. Separation of Concerns

In der Software-Technik hat sich das Prinzip des *Separation of Concerns* [3] für die Entwicklung von Software-Systemen sehr stark bewährt. Mit diesem Prinzip wird die getrennte Entwicklung von Systemteilen erst möglich. Ein bekanntes Beispiel ist die Trennung der Programmierung der Benutzeroberfläche von der Programmlogik und Datenhaltung in einer mehrschichtigen Software-Architektur. Neben der Schichtenbildung gibt es weitere Techniken zur Modularisierung und zur Bildung expliziter Schnittstellen zwischen Software-Komponenten und zur Trennung von technischen und applikationsspezifischen Softwareteilen. Die Modularisierung von Software hat dabei folgende Vorteile:

- Ein Modul lässt sich mit einer adäquaten Teststrategie ausreichend testen.
- Die Schnittstellen unterschiedlicher Systemteile werden explizit beschrieben und gegebenenfalls notwendige Konvertierungen von Werten oder zwischen Programmiersprachen dadurch klarer.
- Die einzelnen Module lassen sich getrennt voneinander entwickeln und gegebenenfalls durch neuere Module austauschen. Sogar das dynamische Nachladen an einer Software-Tankstelle oder zumindest in der Werkstatt kann dadurch modularer vorgenommen werden.
- Die Systemteile lassen sich jeweils mit einer für das Problem und die Entwickler geeigneten Programmiersprache umsetzen. Gerade intelligente Software benötigt oft dafür besonders geeignete programmiersprachliche Konstrukte. Derartige Sprachen sind aber im Allgemeinen weder zertifiziert noch deren vollständig echtzeitfähig.

> Deshalb sind intelligente Funktionen und ihre Absicherung typischerweise in unterschiedlichen Sprachen realisiert.

Die Trennung der Programmteile empfiehlt sich auch bei der Entwicklung von Software im Automobilbau. Dabei sollten Subsysteme gruppiert werden nach:

- harten Realzeitanforderungen und genutzten kontinuierlichen Beschreibungsformen (typisch: Matlab, Simulink),
- diskreten, meist event-basierten Steuerungsaufgaben (Stateflow, Statecharts, C++),
- Nutzung komplexerer Datenstrukturen (C, C++)
- probabilistischen und vor allem adaptiven Berechnungsverfahren (C++, ggf. sogar Prolog o.ae.).

**4. Qualitätssicherung durch Testen**

Im Bereich des Software Engineering von Geschäftsanwendungen sind mittlerweile testorientierte Entwicklungsprozesse etabliert [11], die automatisiert wiederholbare Tests auf allen Ebenen einfordern und von Reviews und Inspektionen flankiert werden. Demgegenüber sind auf dem Gebiet der eingebetteten Software speziell im Rahmen der Integration heterogener Systeme für Automobile solche Tests kaum etabliert. Diese werden immer noch im Wesentlichen durch den manuellen und aufwändigen Test im fahrenden Auto realisiert. Der heutzutage anzutreffende Kundenmehrwert wird immer häufiger durch verteilte Software-Funktionen auf baulich getrennten Hardware-Umgebungen realisiert. Eine integrierte Qualitätssicherung wird aufgrund der speziellen Zulieferer-Situation der Automobilbranche deutlich erschwert. Eine frühe integrierte Qualitätsprüfung der Software von allen Zulieferern würde bereits zur Entwicklungszeit des Fahrzeugs im Rahmen von Prototypen Schwächen der Software selbst, als auch Schwierigkeiten im Integrationsprozess verschiedener Software-Bausteine aufzeigen. Software-in-the-Loop-Maßnahmen zur Qualitätssicherung werden hierbei bislang im Bereich von Reglern eingesetzt, jedoch sind die Modellanforderungen komplex, so dass eine vollständige Simulation der Realität derzeit noch nicht adäquat möglich ist [12]. Die Initiative AUTOSAR [15] will zwar eine Conformance Test Suite zur Einhaltung der definierten Schnittstellen bereitstellen, geht jedoch explizit nicht auf Prozesse zur Qualitätssicherung ein, so dass Conformance-Probleme in der Praxis oft erst spät erkannt werden dürften.

Wie oben beschrieben werden für Steuergeräte-Software oft unterschiedliche Entwicklungswerkzeuge und -techniken kombiniert, so dass die Integrations-Fähigkeit der Module speziell gesichert werden muss [12]. Zur Erreichung eines vorgegebenen Qualitätslevels, der anhand quantitativ überprüfbarer Kriterien zu validieren ist, sind neben den konstruktiven

Maßnahmen durch Architekturmuster und analytischer Berechnungen auf Basis einer Architektur vor allem methodische Regelungvorgaben einzusetzen. Zur Sicherstellung der Einhaltung einer definierten und freigegebenen Spezifikation einer Software können sowohl auf unterschiedlichen Ebenen einer Software-Architektur als auch zu unterschiedlichen Entwicklungszeitpunkten verschiedene Arten von Maßnahmen zur Sicherstellung einer geforderten Software-Qualität getroffen werden. Eine Auswahl dieser Methoden wird in diesem Abschnitt dargelegt. Für eine weiterführende Diskussion sei hier auf [3], [11], [12] und [13] verwiesen.

Für den Nachweis der in der Spezifikation vereinbarten Funktionalität und der möglichst frühzeitigen Erkennung von Fehlern innerhalb der Software werden in modernen Software-Entwicklungsprozessen sowohl automatisierte als auch manuelle Testfälle [11] eingesetzt. Testfälle enthalten in unterschiedlichem Umfang Anwendungsszenarien auf verschiedenen Schichten einer Software. Die Anwendungsszenarien zur Validierung einer Spezifikation können dabei auf Ebene von Methodenaufrufen, die als *Unit-Tests* bezeichnet und in derselben Programmiersprache wie die zu testende Software formuliert werden, über Prüfungen der Interaktionen verschiedener Komponenten im Rahmen von *Integrationstests*, bis hin zu *Akzeptanztests* beim eigentlichen Kunden implementiert werden. Es werden dabei „White-Box-Tests", die detaillierte Kenntnisse der zu testenden Teilkomponente voraussetzen, über „Grey-Box-Tests" bis hin zu „Black-Box-Tests", die von den internen Abläufen nahezu vollständig abstrahieren, unterschieden.

### 4.1. Testen intelligenter Software

Intelligente, adaptive Softwarekomponenten sind grundsätzlich ebenfalls Test-Verfahren zu unterziehen. Automatisierte Tests sind bei hochgradig adaptiver Software besonders zu empfehlen, da so verschiedene Lernsituationen und gelernte Ausgangssituationen bereits offline getestet werden können. Für den Vergleich des erhaltenen Ist-Ergebnisses mit einem gewünschten Soll-Ergebnis kann grundsätzlich verfahren werden, wie oben bei der Einführung von Redundanz beschrieben. Auch hier bleibt allerdings das Problem, dass Ist- und Soll-Ergebnis sich im Rahmen einer gewissen, messbaren Bandbreite unterscheiden dürfen und diese Bandbreite bei komplexen Parametern und Verhalten schwer bestimmbar ist.

Zur Unterstützung des Entwicklers existieren zahlreiche Programme, mit denen häufig wiederkehrende Prozessschritte automatisiert werden können [11], [13]. Werkzeuge dieser Art bilden die technische Grundlage zur fortwährenden und unbeaufsichtigten Überprüfung der einzelnen Quellcodes im Rahmen einer *continuous integration*, um zeitnah auch kumulierte Aussagen über die Qualität der vorliegenden Programme treffen zu können [14].

Zur Vermeidung von überflüssigen Testfällen bietet sich als Ergänzung der reinen Unit-Tests der Einsatz so genannter Überdeckungsmetriken an. Diese berechnen die Anzahl *ausgeführter* Anweisungen in Relation zu allen *ausführbaren* Anweisungen bzw. Kontrollflusspfaden innerhalb einer Methode bei Anwendung eines Testfalls. Darüber lassen sich zum Beispiel Aussagen über fehlende Testfälle als Metrik zur Beurteilung besonders testwürdiger Software-Teile ableiten.

Der Einsatz intelligenter Systeme im Fahrzeug, die Entscheidungen sowohl situationsbezogen im Rahmen eines Entscheidungsfindungsprozesses treffen, als auch gegebenenfalls ein nicht-deterministisches Verhalten aufweisen, ist die effektive Planung von Testfällen unerlässlich für deren Absicherung. Ziel hierbei ist es, aus kontinuierlichen Entscheidungsprozessen diskrete Ablauf-Kategorien zu klassifizieren, die im Rahmen von Testfällen einerseits überprüfbar sind und andererseits in ihrer Vereinigung das intelligente System als solches möglichst vollständig testen. Solche Klassenbildung ist hochgradig abhängig von der zu testenden Funktionalität. Bei komplexen Parametern intelligenter Software sind oft sogar eigenständige Überdeckungsmetriken zu identifizieren, um so die Abdeckung der adaptiven Parameter zu messen.

### 4.2. Eine Testmethodik

Eine mögliche Vorgehensweise zur Validierung von adaptiver Software mit kontinuierlichen und diskreten Anteilen kann wie folgt skizziert werden: (1) Bestimmung der adaptiven Software-Komponenten, die entweder kontinuierliche Datenströme oder Kondensate historischer Daten zur Entscheidungsfindung nutzen. (2) Untersuchung dieser Komponenten auf wiederkehrende Verhaltens-Muster, anhand derer eine Entscheidung getroffen wird. (3) Umsetzung ausreichend vieler automatisierter Testfälle zur Überdeckung der Muster.

Ein Beispiel für eine solche Komponente ist etwa die Entscheidungsfindung zur Einleitung eines Überholmanövers auf einer Autobahn. Ein Überholmanövers besteht im Wesentlichen aus dem Ausscheren auf die benachbarte Fahrspur, der Vorbeifahrt und das Wiedereinscheren auf die ursprüngliche Fahrspur. Im kontinuierlichen Entscheidungsprozess müssen über mehrere Betrachtungszeitpunkte in Folge eine Reihe von Vorbedingungen erfüllt sein, bevor ein Überholmanöver eingeleitet werden darf: beispielsweise darf die benachbarte Fahrspur nicht belegt sein bzw. auf dieser Spur das erste Fahrzeug hinter dem Ego-Fahrzeug darf während einer situationsabhängigen Zeitdauer $\Delta t$ nicht dieselbe Höhe wie das eigene Fahrzeug erreichen können. Hier lässt sich erkennen, dass komplexe Entscheidungsprozesse oft in einfachere Entscheidungskomponenten zerlegt werden können.

## 5. Zusammenfassung

Ziel einer guten Software-Architektur ist zunächst die Zerlegung komplexer Software-Komponenten. In einem von Beginn an umgesetzten integrativen Testprozesses werden die korrekte Reihenfolge und die Abhängigkeiten der einzelnen Software-Komponenten über Testfälle sichergestellt. Die Parametrisierung adaptiv lernender Software-Komponenten wird dabei ganz besonders intensiv getestet und auf Abweichungen vom Norm-Verhalten untersucht. Durch ein Bündel an softwaretechnischen Maßnahmen werden so einerseits die schnelle Entwicklung und Qualitätssicherung intelligenter Software für das Auto und andererseits die softwarebasierte Absicherung im Fahrbetrieb ermöglicht.